\documentclass{hep99}

\usepackage{epsfig}
\begin{document}

\title{An investigation on the $b$-quark mass}

\author{Fabrizio Palla$^1$\dag}
%

\address{$^1$ Istituto Nazionale di Fisica Nucleare \\
Sezione di Pisa, Via Livornese 1291 \\
I-56010 San Piero a Grado (PI) Italy}

\abstract{In this article I will review the measurements of the $b$ quark mass performed by ALEPH and DELPHI. A large set of observables has been used together with detailed studies on jet algorithms. Very clear effects due to the $b$ quark mass running are observed , even if a wide spread in the results and the large systematic errors coming from hadronization corrections prevents to extract an average value of the running $b$ quark mass.}

\maketitle

\fntext{\dag}{E-mail: {\tt Fabrizio.Palla@cern.ch}}

\section{Introduction}
The $b$ quark mass is one of the fundamental parameters of the QCD Lagrangian.
However, due to confinement, quarks do not appear as free particles and therefore the definition of their mass is ambiguous. In fact, quark masses can be either defined as for free particles (such as the leptons) as the position of the pole of the propagator, or they can be interpreted as running coupling constants in the Lagrangian. In the former definition the mass is called ``pole mass'' and does not run with energy; in the latter the mass is called ``running mass'' and it can run if measured at different scales.

Most of the $b$ quark mass determinations have been performed at rather low scales \cite{PDG}. It is therefore interesting to measure this parameter at higher scales such as the one offered by LEP.

At LEP, the possibility to test the running of the $b$ quark mass within the framework of the perturbative QCD has been not considered until very recently. The reason is that the effects of the mass become rapidly very small with increasing energy for many observables, since they are proportional to $m_b^2/M_Z^2$ ($\cal O$ (0.1\%)) for instance as in the total cross section. For other specific quantities, such as jet rates, the effects are of the order of few percent  due to the fact that the energy scale is lowered to $y_{cut}\times M_Z^2$, where $y_{cut} \approx 0.01$ is the jet resolution parameter.

The measurements are performed by measuring the ratio of observables in $b$ and $uds$ induced events.

In this article I will review the measurements done by ALEPH\cite{ALEPH_paper} using a large set of event shape observables and by DELPHI\cite{DELPHI_paper1,DELPHI_paper2} using the three-jet rate. The different  sensitivity to the hadronization corrections and next-to-leading order corrections can be used to investigate the interpretation of the results.

\section{Analysis Method}

The method to extract the $b$ quark mass is based on measuring the ratio $R$ of any infrared safe observable $O$ computed for  $b$ and $uds$ induced events and assuming $\alpha_S$ universality:
\begin{equation}
R_{b/uds} = \frac{O_b}{O_{uds}}.
\end{equation}

As observables the following variables have been studied by ALEPH:
\begin{itemize}
\item The rate of the three-jet events, where jets are reconstructed using the Durham clustering algorithm using $y_{cut}=0.02$.
\item The first and the second moments of the event shape variables Thrust T, C-parameter C, $y_3$, Total and Wide Jet Broadening, $B_T$ and $B_W$.
\end{itemize}

DELPHI has used the three and the four jet events rate. 


Firstly events are tagged as to originate from $b$ or $uds$ quarks. Both
ALEPH and DELPHI use a $b$ tag based on lifetime and mass information
\cite{btags}. In order to tag $uds$ events ALEPH uses a dedicated tag using
lifetime  as well as kinematic information  combined to a discriminant
variable. DELPHI uses instead a different region of the
lifetime tag where the majority of the events are coming from $uds$.
Typical $b$ quark purities for the $b$ enriched sample are of the order
of about 85\% and  efficiencies of about 85\% for ALEPH and 50\% for DELPHI.
The $uds$ purity is about 80\% with 60\% efficiency for both.

The ratio is then corrected for hadronization, detector and tag biases.

The $b$ quark mass is extracted by comparing the corrected measured ratio $R_{b/uds}^P$ with the predictions for each observable.

The general form of the NLO prediction for $R_{b/uds}^P$ as a function of
the $b$ quark mass is of the form
\begin{equation}
R_{b/uds}^P = 1+\frac{m_b^2}{M_Z^2}\left[ b_0(m_b) + \frac{\alpha_S}{2\pi}b_1(m_b)\right]
\end{equation}

The coefficient functions $b_0$ and $b_1$ have been computed for the three
and four jet rate ratio \cite{Rodrigo,zbb4}. The three jet
rate ratio allows a NLO prediction, while for the four jet rate only a LO
prediction is available. For the other variables used by ALEPH the
predictions have been obtained using the MC generators ZBB4\cite{zbb4}
and EVENT\cite{event} which are correct to NLO.
The hadronization corrections have been evaluated for all variables by
computing the relevant observables at parton and at hadron level.

\section{ALEPH Analysis}

As shown in table \ref{tab_had_aleph} for the majority of the observables
the hadronization corrections are sizeable.
Therefore it has been decided to extract the
$b$ quark mass from the ones which receive corrections smaller than 10\%:
these are the three jet rate ratio and the second moment of the Wide Jet
Broadening $BW_2$. The first two moments of the $y_3$ variable being very much
correlated to the three jet rate have not been used for the final result.
Nonetheless the mass has been computed for all variables.
The results are given in table \ref{tab_had_aleph}, where the error is
statistical only. As it can be seen there
is a large scatter in the results. It is worth noticing
that the result extracted from the first moment of the $y_3$ are quite
different from the one extracted from the three jet rate ratio, despite of the
fact that the two variables are very much correlated. This might indicate that
there are still large uncontrolled biases from hadronization and/or that the
NNLO corrections are sizeable.

Systematic uncertainties can be divided in three categories: the ones coming
from uncertainties on the tagging biases, from the hadronization and from
the missing higher order corrections.
The first ones are evaluated by varying the gluon splitting rate which
influences the purity of the lifetime tag and by varying the cut variables
so that the effect of the tag corrections are twice the effect of
the mass on a given variable. This results in varying the tag purities by about 5\%.
The uncertainty coming from hadronization correction has been evaluated mainly
varying the $Q_0$ Jetset parameter from 1 to 4 GeV and taking half of the
difference between Jetset and Herwig predictions.
The uncertainty due to missing higher orders has been evaluated by varying the
scale from 0.1 $M_Z$ to 2 $M_Z$ and by the difference in the result in the
value of the running $b$ quark mass directly extracted and the one 
extracted from the following procedure: from the pole mass measurement one
translates it to a running mass at the scale of the $b$ quark mass $m_b(m_b)$
and finally to the $Z$ mass scale $\bar m_b(M_z)$.

The results for the running $b$ quark mass as extracted using the three jet rate
ratio and the second moment of the Wide Jet Broadening are respectively:
\begin{eqnarray}
m_b(M_Z) =& 3.04^{+0.37}_{-0.34}(stat)^{+0.44}_{-0.39}(syst)\nonumber\\
         &^{+0.72}_{-0.59}(hadr)^{+0.20}_{-0.42}(theo) GeV/c^2. \\
m_b(M_Z) =& 3.78\pm 0.14(stat)\pm 0.17(syst)\nonumber\\
         &\pm 0.10 (hadr)^{+0.12}_{-0.13}(theo) GeV/c^2.
\end{eqnarray}

\begin{table}[htb]
  \begin{center}
    \begin{tabular}{|c|c|c|}
      \hline
      \hline
      $O$ & $H_{b/l}$& $m_b(M_Z)$ \\
      \hline
      \hline
      $T_1$ & 1.142 & $4.48\pm0.09$ \\

      $T_2$ & 1.139 & $4.84\pm0.20$ \\

      $C_1$ & 1.175 & $4.41\pm0.06$ \\

      $C_2$ & 1.181 & $4.69\pm0.12$ \\

  $y_{3_1}$ & 1.029 & $3.89\pm0.28$ \\

  $y_{3_2}$ & 0.990 & $3.51^{+1.50}_{-0.95}$ \\

  $B_{T_1}$ & 1.302 & $3.94\pm0.03$ \\

  $B_{T_2}$ & 1.333 & $3.57\pm0.06$ \\

  $B_{W_1}$ & 1.142 & $4.74\pm0.05$ \\

  $B_{W_2}$ & 1.093 & $3.78\pm0.14$ \\

 $R^{Dur}_{3j}(0.02)$ & 0.989& $3.04^{+0.37}_{-0.34}$ \\

    \hline
    \hline
    \end{tabular}
    \caption{Table of the ratio of hadronization corrections $H_{b/l}$,   
           and the extracted running $b$-quark mass $m_b(M_Z)$. The errors are
             statistical only. 
             \label{tab_had_aleph}} 
  \end{center}
\end{table}

\section{DELPHI Analysis}

The first measurement of the running $b$ quark mass at scales of the $Z$ mass
has been performed by DELPHI, using 1992 to 1994 data and three jet rate ratio
using the Durham algorithm for jet clustering.\cite{DELPHI_paper1}
A new analysis has been performed on 1994 and 1995 data using both Durham and
Cambridge jet algorithms, in addition to a better b-tag algorithm. The Cambridge algorithm has been proven to have
smaller NLO corrections as well as a reasonable stable hadronization correction
as a function of $y_{cut}$ down to very low values.

\begin{figure}
\begin{center}
\epsfig{file=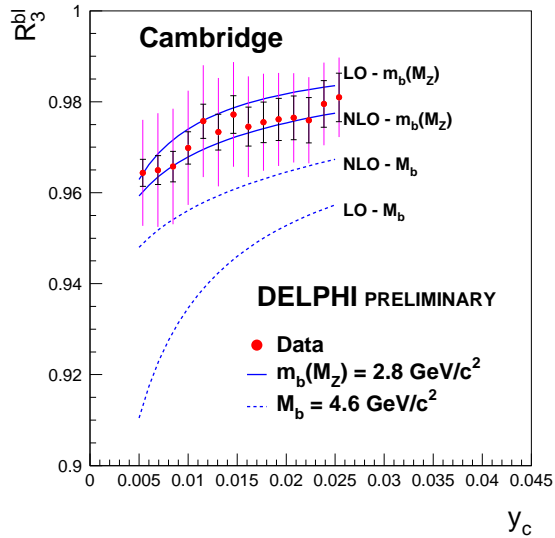,width=8cm}
\end{center}

\caption{Corrected data values for the three jet ratio $R_3^{bl}$ using the Cambridge algorithm as a function of the $y_{cut}$ compared with the theoretical predictions at LO and NLO in terms of the pole mass $M_b$ (dashed lines) and in terms of the running mass $m_b$(solid lines).\label{fig_delphi}}
\end{figure}

The same systematic sources have been considered as ALEPH. Some differences
arise in the way they are evaluated. The ones coming form the tags have been
evaluated by changing the tag purities by 1\% and taking the difference between the
new and old tag. The hadronization corrections have been evaluated by changing
the fragmentation tuning of the Monte Carlo by 2 sigma and by comparing Jetset
with Herwig. The theoretical uncertainty due to missing higher orders has been
evaluated by changing the scale from  0.5 $M_Z$ to 2 $M_Z$ and by comparing the
results obtained by running the pole $b$ mass to $\bar m_b(M_Z)$ as explained before.
The analysis performed using the Durham algorithm gives consistent results
with the old data analysis. 
The values of the three jet rate ratio using the Cambridge algorithm are shown in figure \ref{fig_delphi}.  The data are not compatible with the prediction using a pole mass of 4.6 GeV/c$^2$.

The result obtained  with the Cambridge algorithm at $y_{cut}=0.005$ for the three jet rate ratio is
\begin{eqnarray}
  m_b(M_Z)  =& 2.61 \pm 0.18 (stat) ^{+0.45}_{-0.49} (hadr) \nonumber \\
             & \pm 0.04 (tag)   \pm 0.07 (theo) GeV/c^2.
\end{eqnarray}
where the average between the direct extraction of the running mass  $m_b(M_Z)$ and the one extracted from the running of the pole mass  $\bar m_b(M_Z)$.  

Despite of the small dependence of the measured mass from the value of the 
jet resolution parameter $y_{cut}$ (changing $y_{cut}$ from 0.005 to 0.025 the value of $m_b(M_Z)$ changes by 0.15 GeV/c$^2$), the result still suffers from large hadronization corrections.

Effects of the running of the $b$ quark mass have also been measured in the four jet rate ratio as reported by \cite{Garcia}.

\section{Conclusions}

 ALEPH has studied a large set of observables to extract the running $b$ quark mass. A very nice agreement has been with the DELPHI results using a three jet rate ratio. A wide spread in the  $b$ quark masses extracted from all the various observables might indicate uncontrolled biases from hadronization and/or NNLO corrections.

DELPHI has measured the running $b$ quark mass using the Cambridge jet clustering algorithm finding a very good agreement with NLO calculations and small theoretical errors, but still suffers from large hadronization systematics.

Due to these findings the interpretation of the results have to be taken with care and do not allow to extract an average  of the running $b$ quark mass from
different observables.

\end{document}